\documentstyle[prl,aps,multicol,epsf]{revtex}
\begin{document} 
\draft 
\title{Doping dependent quasiparticle band structure in cuprate
superconductors}

\author{R. Eder$^1$ , Y. Ohta$^2$, and G. A. Sawatzky$^1$}
\address{$^1$Department of Applied and Solid State Physics, 
University of Groningen,
9747 AG Groningen, The Netherlands\\ 
$^2$Department of Physics, Chiba University, Chiba 263, 
Japan} 
\date{\today}
\maketitle

\begin{abstract} 
We present an exact diagonalization study of the 
single particle spectral function in the so-called
$t$$-$$t'$$-$$t''$$-$$J$ model in $2D$. As a key result, we find that
hole doping leads to a major reconstruction of the quasiparticle
band structure near $(\pi,0)$:
whereas for the undoped system the quasiparticle
states near $(\pi,0)$ are deep below the top of the band near 
$(\pi/2,\pi/2)$,
hole doping shifts these states up to $E_F$, resulting in extended
flat band regions close to $E_F$ and around $(\pi,0)$. 
\end{abstract} 
\begin{multicols}{2}

\begin{figure}
\epsfxsize=9cm
\vspace{-0.75cm}
\hspace{-0.5cm}\epsffile{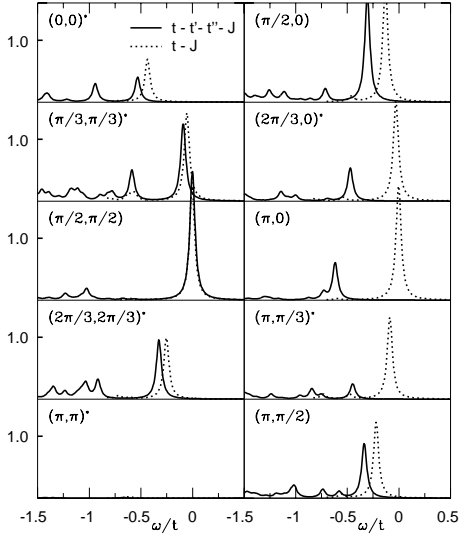}
\vspace{-1.0cm}
\narrowtext
\caption[]{PES spectra at half-filling for
the $16$ and $18$ site cluster.
Momenta marked by an asterisk refer to the $18$ site cluster,
$\delta$-peaks are replaced by Lorentzians of width $0.03t$. The
top of the band for the $4\times 4$ cluster
is the zero of energy, the spectra for $18$ sites
are shifted uniformly to produce a smooth dispersion.}
\label{fig1}
\end{figure}

\begin{figure}
\epsfxsize=10cm
\vspace{-0.75cm}
\hspace{-0.5cm}\epsffile{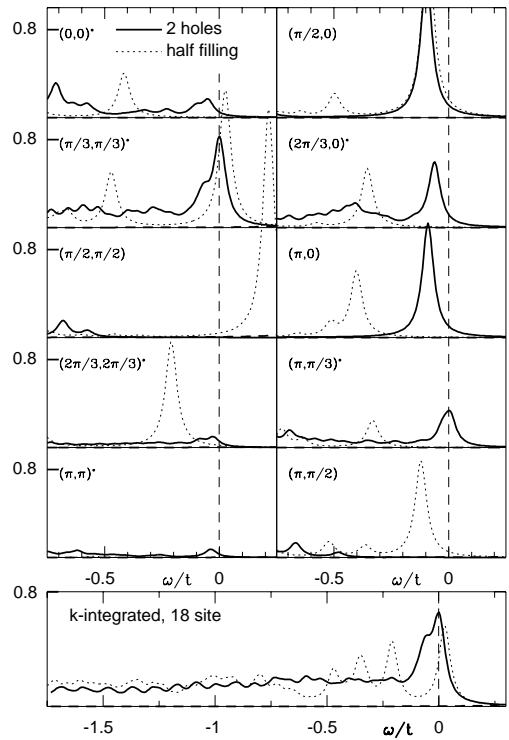}
\vspace{0.25cm}
\narrowtext
\caption[]{
Doping dependence of PES spectra for the
$16$ and $18$ site cluster. The chemical potential $\mu_-$
for $18$ sites is the
zero of energy, the spectra for $4\times 4$ are shifted uniformly
to give a smooth dispersion of the `quasiparticle peak'.}
\label{fig2}
\end{figure}

\begin{figure}
\epsfxsize=9cm
\vspace{-1.5cm}
\hspace{-0.5cm}\epsffile{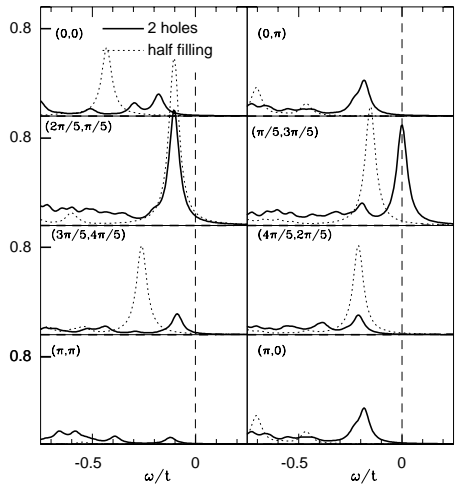}
\vspace{-1.0cm}
\narrowtext
\caption[]{As Figure \ref{fig2}, but for the $20$-site cluster. 
The zero of energy is $\mu_-$.}
\label{fig3}
\end{figure}

\begin{figure}
\epsfxsize=9cm
\vspace{-0.5cm}
\hspace{-0.5cm}\epsffile{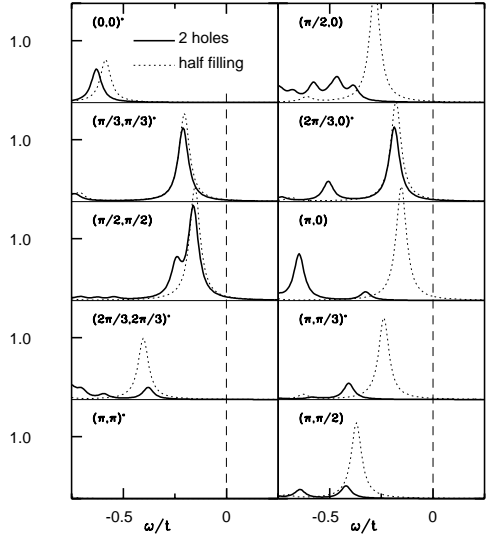}
\vspace{-1.0cm}
\narrowtext
\caption[]{Doping dependence of PES spectra for the
$16$ and $18$ site cluster for the $t$$-$$J$ model.}
\label{fig4}
\end{figure}

Photoemission experiments on the
insulating Heisenberg antiferromagnet Sr$_2$CuO$_2$Cl$_2$\cite{Wells} 
and the underdoped cuprate superconductor
Bi$_2$Sr$_2$CaCu$_2$O$_{8+\delta}$\cite{Marshall}
have provided unexpected results.
The experimentally measured quasiparticle dispersion in Sr$_2$CuO$_2$Cl$_2$
shows a band maximum at momentum $(\pi/2,\pi/2)$ with a relatively
isotropic effective mass, the first ionization states
for momentum $(\pi,0)$ are found at an energy of $\approx 0.3eV$
below the top of the band. This `gap' at $(\pi,0)$
then seems to gradually close with doping: ARPES spectra
for the underdoped compound Bi$_2$Sr$_2$CaCu$_2$O$_{8+\delta}$\cite{Marshall}
show that the parts of the band structure
near $(\pi,0)$ shift to a binding 
energy between $0.2$ and $0.15eV$ as the hole concentration increases
and for the optimally doped compounds the quasiparticle band
near $(\pi,0)$ is within $50meV$ below the chemical potential,
forming extended flat regions\cite{Dessau}.
Taken together, and assuming that the `effective Hamiltonians' in the
different compounds do not differ drastically, these data
suggest a continuous but quite
substantial deformation of the quasiparticle band
structure with doping, resulting essentially in a shift of the
parts near $(\pi,0)$ towards lower binding energies.
This scenario poses a formidable problem to theory
in that no numerical study of either $t$$-$$J$ model\cite{Shimozato}
or Hubbard model\cite{Bulut,Preuss,Eskes}
has ever shown a behavior which is even remotely comparable:
for both models, the quasiparticle band is found immediately
below $E_F$ at half-filling and stays there more or less
independently of doping.
Here it should be noted that the experimentally measured
binding energy of the quasiparticle states
near $(\pi,0)$ in the underdoped
compounds\cite{Marshall} is quite large
on the scale of the quasiparticle bands 
in the strong correlation models, which is $\sim 2J$$=$$260\;meV$.
It is therefore not plausible that this deformation
should have been simply overlooked due to finite size or
temperature effects.\\
For the undoped models the only way known so far to reproduce
the band structure of 
Sr$_2$CuO$_2$Cl$_2$ within the framework of
$t$$-$$J$ like models is the inclusion of longer range hopping 
terms\cite{Nazarenko}. To obtain a good fit to experiment requires
quite substantial hopping integrals between both second and third nearest
neighbors, resulting in the now intensively studied 
$t$$-$$t'$$-$$t''$$-$$J$ model. 
Anticipating that the microscopic model parameters, i.e.
in particular $t'$ and $t''$, are a generic feature of the CuO$_2$ plane,
this seems to suggest that a description of ARPES experiments in terms 
of strong correlation models would require a very strong doping dependence
of these parameters. In this work we present
numerically exact results for small clusters which show 
that on the contrary the $t$$-$$t'$$-$$t''$$-$$J$ model with 
fixed, doping independent 
parameters can account for the experimentally observed doping dependence
of the quasiparticle band structure. We find that upon doping, the states 
near $(\pi,0)$ shift towards the chemical potential so that there is
a strongly doping dependent quasiparticle band
structure and in particular an extended flat band region around $(\pi,0)$
for nearly optimal doping.\\
We study the $t$$-$$J$ model with
hopping integrals $t$ between nearest
neighbors, $t'$ between $2^{nd}$ nearest (i.e. $(1,1)$-like) neighbors
and $t''$ between $3^{rd}$ nearest (i.e. $(2,0)$-like) neighbors. 
We retain only the exchange integral $J$ between nearest neighbors.
We choose $t$ as the 
unit of energy and adopt the values $J$$=$$0.3t$ , $t'$$=$$-0.35t$ and 
$t''$$=$$0.25t$. These parameters are within the 
range of values derived from the three band model
although there is some variation between different estimates\cite{Feiner};
in fact our main justification is the good
fit to the experimental quasiparticle dispersion in
Sr$_2$CuO$_2$Cl$_2$ which they provide (see below). 
We have moreover verified that our key result, namely the doping induced
change of the quasiparticle dispersion, is not sensitive to this
choice, and in particular occurs also for the case $t''$$=$$0$.
The quantity of interest is the photoemission (PES) spectrum
\[
A_{PES} (\bbox{k},\omega) = \Im
\langle \Psi_0 | c_{\bbox{k},\sigma}^\dagger 
\frac{1}{\omega + ( H - E_0 ) - i0^+}
c_{\bbox{k},\sigma}
|\Psi_0\rangle,
\]
where $|\Psi_0\rangle$ and $E_0$ denote the ground state
wave function and energy.
We have evaluated this function for the standard $16$, $18$
and $20$ site clusters by means of the Lanczos algorithm\cite{Dagoreview}. 
To begin with, Figure \ref{fig1}
shows the PES spectrum at half-filling, combining
results for $16$ and $18$ sites, for both the
full $t$$-$$t'$$-$$t''$$-$$J$ model and for the case
$t'$$=$$t''$$=0$.
We consider only the first ionization states,
within a few $J$ from the top of the band.
As discussed by various authors\cite{Nazarenko,Sushkov,Chernychev}, 
the $t'$ and $t''$ terms lift
the (quasi) degeneracy of the quasiparticle states at $(\pi/2,\pi/2)$
and $(\pi,0)$ which occurs in the $t$$-$$J$ model, 
\noindent 
shifting $(\pi,0)$ to substantially more negative 
(on the scale of the total band width)
binding energy.
Another notable effect is the change
in the weight of the quasiparticle peaks as compared to
the pure $t$$-$$J$ model: whereas the peak at
$(\pi/2,\pi/2)$ retains its original weight,
the peaks near $(\pi,0)$, which
had a comparable weight in the pure $t$$-$$J$ model, become quite
weak\cite{Sushkov}. This may well correspond to the experimental
situation, where a quasiparticle peak near $(\pi,0)$ is not actually
resolved\cite{Wells}. All in all, the spectra at half-filling allow for 
rather direct comparison with the experimental results.\\
We proceed to the hole doped case. Table \ref{tab1}
summarizes some properties of the two-hole ground states.
For the $16$-site cluster the absolute ground state
has finite momentum $(\pi,0)$. We believe that the existence of this
state is an artefact of the special geometry
of the $4\times4$ cluster, which in the case $t'$$=t''$$=0$ 
is known\cite{HasegawaPoilblanc}
to lead to spurious ground state degeneracies; in fact, no other cluster
does have a finite momentum ground state for even hole number.
The PES spectra for this state
show a strong
asymmetry between momenta along $(1,0)$ and $(0,1)$, which also
indicates that this state is not representative for a bulk system.
Finally,
the lowest two hole state with zero momentum in the $16$ site cluster
(also listed in Table \ref{tab1})
is nearly
degenerate with the $(\pi,0)$ state, the separation in energy/site being 
in fact much smaller
than the variation of this quantity between different clusters.
We therefore believe that it is reasonable to
dismiss the $(\pi,0)$ state
and consider instead the lowest zero momentum state.
Next, the $20$ site cluster
is special
in that already its {\em half-filled} ground state
has $d_{x^2-y^2}$ symmetry (whereas it is $s$ for all other
clusters). Adding two holes
does not change the ground state symmetry (see Table \ref{tab1}), 
so that the hole pair obviously
corresponds to an object with $s$-like symmetry.
Thus, for all available clusters the
symmetry of the zero momentum ground state of a hole pair is changed from
$d_{x^2-y^2}$ for the $t$$-$$J$ model to either
$s$ or $p$-like symmetry. More detailed
analysis shows that the singlet $s$ and triplet $p$
states are nearly degenerate
in all clusters, so that subtle geometrical features of the
different clusters decide which one is the actual ground state.\\
After these preliminaries, we proceed to the
discussion of the spectral function.
Figure \ref{fig2} 
shows the photoemission spectrum for
the $2$ hole ground states of $16$ and $18$-site 
\noindent clusters,
Figure \ref{fig3} presents analogouos data for $20$ sites.
The spectra for the $18$ site cluster have been calculated
for the $p_x$-like ground state with $S_z$$=$$0$. 
For momenta along $(1,1)$, the
$p_y$-like state gives identical spectra by symmetry;
the spectra at $(2\pi/3,0)$ and $(\pi,\pi/3)$ in principle could
be different, but numerical evaluation shows that they differ only 
negligibly. We have chosen the zero of energy at the
excitation energy of the first ionization state, which corresponds
to setting the chemical potential equal to the `left side derivative'
$\mu_-$$=$$E_0(N)-E_0(N-1)$. Assuming that the quasiparticle peaks
in the cluster spectra approximately interpolate the dispersion
for the thermodynamic limit, this estimate for the chemical
potential may be expected to be accurate up to the
`discretization error'
$v_F | \bbox{k}_F^c - \bbox{k}_F|$ with $v_F$ is the
Fermi velocity, $\bbox{k}_F^c$ the `Fermi momentum' of the cluster
(i.e. the momentum of the first ionization state), and
$\bbox{k}_F$ the true Fermi momentum in the infinite system which is 
closest to $\bbox{k}_F^c$. This error stems from the
discreteness of the $\bbox{k}$-mesh and is unavoidable in any finite cluster.
One may expect, however, that it is a small fraction of the
band width, and thus not really appreciable.
Comparison with the spectra at half-filling then reveals
quite a substantial 
change of 
\noindent the band 
structure seen in photoemission:
whereas e.g. the topmost PES peaks at $(\pi/2,0)$ and $(\pi,0)$
in the $16$-site cluster
were separated by an energy of $\approx 0.3t$ at half-filling
they are now practically at the same energy.
Similarly, the topmost PES peaks at $(2\pi/3,0)$ and
$(\pi,\pi/3)$ were $\approx 0.4t$ below
the intense peak at $(\pi/3,\pi/3)$ in the $18$-site
cluster at half-filling, but are practically
at the same energy for the doped case.
In the $20$-site cluster
the topmost PES peaks at $(2\pi/5,\pi/5)$ and
$(\pi,0)$ differed by $\approx 0.4t$ in the undoped case
but are again nearly degenerate for the two hole ground state.
Similarly, the PES peak at $(\pi/5,3\pi/5)$ is approximately
degenerate with $(2\pi/5,\pi/5)$ for half-filling, but
is appreciably shifted upwards relative to this peak in the spectra
for the doped ground state. Irrespective of the cluster geometry
and ground state quantum numbers the spectra for the doped case
thus show a shift of the peaks near $(\pi,0)$ towards
lower binding energies so that they become (degenerate with) 
the first ionization states. 
This shift in energy is usually accompanied
by an increase in peak intensity. This is most clearly seen in the
$16$ site cluster, but also for the $18$ and $20$ site clusters
there is a slight increase in the intensity of the peaks themselves
and also an upward shift of incoherent weight. 
For other momenta, such as $(0,0)$ or $(2\pi/3,2\pi/3)$
a similar upward shift of 
spectral weight can be seen; in these cases
however, the well defined peaks at half-filling 
disappear completely, and the spectra resemble incoherent
continua. The doping induced
deformation of the band structure may have a remarkable consequence,
which is also illustrated in Figure \ref{fig2}: the upward shift of
the quasiparticle band near $(\pi,0)$ 
leads to the seemingly counterintuitive
result that near $E_F$ the decrease of electron density
leads to an {\em increase} of electron removal weight
in the $\bbox{k}$-integrated PES spectrum. It is tempting to compare 
this to experimental results on cuprate superconductors, which also
show an increase of PES weight at $E_F$ with increasing doping\cite{Allen}.
It should be noted, that our sampling of the 
Brillouin zone is rather coarse and finite-size
effects large, so that the $\bbox{k}$-integrated
spectra probably have little quantitative significance.
However, the upward shift of a weakly dispersive part of the
band structure by $0.3-0.4t$$\approx$$0.2eV$ (possibly even accompanied by an 
increase of its spectral weight) is a very plausible 
mechanism for such an
increased PES weight near $E_F$. The up to now mysterious
increase of spectral weight with doping at the top of the
band (the Fermi energy for the metals) thus is quite naturally
explained by this model, showing that the formation of
`new states' in the charge transfer gap with doping
is not required to explain the experiments.\\
The strong doping dependence of the quasiparticle band structure
in the $t$$-$$t'$$-$$t''$$-$$J$ model
becomes very clear in a comparison with the
`pure' $t$$-$$J$ model, shown in Figure \ref{fig4}. When compared with
Figure \ref{fig2}, this figure shows the remarkable `rigidity' of the
band structure: the quasiparticle peaks in the interior of the 
antiferromagnetic Brillouin zone (AFBZ) 
persist with practically unchanged spectral weight
and dispersion, peaks outside the AFBZ loose most of their
weight, but persist as `shadow band'.
For momenta near $(\pi,0)$ the PES peaks cross to the inverse
photoemission spectrum with almost unchanged weight
(the fact that this happens near $(\pi,0)$ rather than 
$(\pi/2,\pi/2)$ is due to the $d_{x^2-y^2}$ symmetry of the
two-hole ground state\cite{Shimozato}).
In cases where there is an
appreciable change of the peak positions, the PES peaks usually
shift away from the chemical potential as compared
to the strong peaks right at $E_F$; moreover, 
the weight of the peaks 
usually decreases. The general trends with doping
therefore are opposite to that of the $t$$-$$t'$$-$$t''$$-$$J$ model.
Moreover, the spectra for the pure $t$$-$$J$ model on the average
look `more coherent', i.e. the decay of sharp peaks into
incoherent continua is much less pronounced as for the 
$t$$-$$t'$$-$$t''$$-$$J$ model.
As far as a `dispersion'
can be assigned, the band for
the $t$$-$$t'$$-$$t''$$-$$J$ model essentially approaches that of the
ordinary $t$$-$$J$ model, i.e.
$t'$ and $t''$ seem to be
renormalized towards 
zero with doping.
This suggests a loss of `spin coherence' as the source of
the doping induced band deformation:
in a N\'eel state, 
\noindent 
transferring a hole to a second- or third-nearest neighbor
gives precisely the same state as if the hole had been created
there in the first place. The $t'$ and $t''$ terms therefore
contribute directly to the coherent motion
of the hole, and consequently have a strong impact on the dispersion.
On the other hand,
when the spin correlation length is comparable
or shorter than the `range' of the
$t'$ and $t''$ terms, a spin transferred
by these may no longer
`fit in' to its new environment, so that the hopping process
with a high probability produces spin excitations
and consequently becomes incoherent. In other words,
rather than contributing to the coherent hole motion,
the $t'$ and $t''$ terms now enhance the emission of
spin excitations. This would also explain
the generally `more incoherent' character of the PES spectra
in the $t$$-$$t'$$-$$t''$$-$$J$ model.
By similar arguments the introduction of frustration at half-filling
should also lead to a deformation of the quasiparticle band
with long distance hopping, as indeed seems to be the case\cite{Hayn}.\\
In summary, we have studied the doping dependence of the
photoemission spectrum for the $t$$-$$t'$$-$$t''$$-$$J$ model.
Using parameter values which give a realistic fit to
the experimental quasiparticle band structure
in the undoped Heisenberg antiferromagnet $Sr_2 Cu_2 Cl_2 O_2$
we found a substantial doping dependence of the
topmost photoemission peaks: upon doping, the quasiparticle band
near $(\pi,0)$ moves to lower binding energy, and becomes degenerate
with the first ionization states. This resuls in particular
in an increase of $\bbox{k}$-integrated spectral weight 
near $E_F$ with doping.
While exact diagonalization does not allow for
a sufficiently smooth variation of the hole concentration
to map out the detailed doping dependence,
the rather substantial deformation of the
quasiparticle band without changing any microscopical
parameter of the model suggests
a unifying point of view for 
both Sr$_2$CuO$_2$Cl$_2$ and cuprate superconductors.
As the mechanism for the doping induced
deformation we propose the interplay between 
longer range hopping and decreasing spin correlation length.
As N\'eel order is lost over the range of the $t'$ and $t''$ terms,
the effect of the latter changes from coherent
hole propagation to emission of spin excitations, leading
to the deformation of the band structure. Taking this further,
one might expect that a degradation of the spin correlations due to
increasing temperature has an analogous effect, i.e. an
overall `flattening' of the band structure. The resulting
increase of the band mass with temperature then could provide
a very simple explanation for much of the `spin gap' behaviour
observed e.g. in the spin susceptibility and entropy
of cuprate superconductors\cite{Loram}.\\
This work was supported by Nederlands Stichting voor Fundamenteel
Onderzoek der Materie (FOM), Stichting Scheikundig Onderzoek
Nederland (SON) and the European Community.

\begin{table}
\narrowtext
\caption{Two hole ground state properties for various cluster
sizes: momentum $\bbox{k}_0$, point group symmetry $R$,
total spin $S_{tot}$
and energy/site $E_0/N$.}
\begin{tabular}{c | c c c c}
 $N$ & $16$ & $16$ & $18$ & $20$ \\
\hline
 $\bbox{k}_0$ & $(0,0)$ & $(\pi,0)$ & $(0,0)$ & $(0,0)$ \\
$R$     &   $s$   &    $-$    & $p$     & $d_{x^2-y^2}$\\
$S_{tot}$     &   $0$   &    $0$    & $1$     & $0$\\
 $E_0/N$      & $-0.6309$ & $-0.6335$ & $-0.6167$ & $-0.5917$ \\
\end{tabular}
\label{tab1}
\end{table}

\end{multicols}
\end{document}